\documentclass[notitlepage,letterpaper, 11pt]{article}
\usepackage[ansinew]{inputenc} 
\usepackage{amsmath}
\usepackage{amsfonts}
\usepackage{amssymb}
\usepackage{graphicx}
\usepackage{subfig}
\usepackage{cancel}
\usepackage{booktabs}
\usepackage{color}
\usepackage{bigints}
\usepackage[top=1cm, bottom=2cm,outer=2cm, inner=2cm]{geometry}
\usepackage{amsthm}
\usepackage{lineno}
\usepackage{tensor}

\begin{document}

\title {Relativisitic non-pascalian fluid as a density contribution} 
\author{
\textbf{Justo Ospino}\thanks{\texttt{j.ospino@usal.es}} \\
\textit{Departamento de Matem\'atica Aplicada and} \\
\textit{Instituto Universitario de F\'isica Fundamental y Matem\'aticas,}  \\ 
\textit{Universidad de Salamanca, Salamanca, Spain;} \\
\textbf{Daniel Su\'arez-Urango}\thanks{\texttt{danielfsu@hotmail.com}}, \\ 
\textit{Escuela de F\'isica, Universidad Industrial de Santander,  }\\ 
\textit{Bucaramanga 680002, Colombia}; \\
\textbf{Laura M. Becerra}\thanks{\texttt{laura.becerra@umayor.cl}} \\
\textit{Centro Multidisciplinario de F\'isica, Vicerrector\'ia de Investigaci\'on,}\\
\textit{Universidad Mayor, Santiago de Chile 8580745, Chile} \\
\textbf{H. Hern\'andez}\thanks{{\tt hector@ula.ve}} and \textbf{Luis A. N\'{u}\~{n}ez}\thanks{\texttt{lnunez@uis.edu.co}} \\
\textit{Escuela de F\'isica, Universidad Industrial de Santander,}\\ 
\textit{Bucaramanga 680002, Colombia} and \\
\textit{Departamento de F\'{\i}sica,} \\
\textit{Universidad de Los Andes, M\'{e}rida 5101, Venezuela.} 
}
\maketitle
\begin{abstract}Understanding the role of pressure anisotropy and dissipation is crucial for modelling compact objects' internal structure and observable properties. In this work, we reinterpret local pressure anisotropy in relativistic stellar structures as an additional contribution to the energy density. This perspective enables the formulation of anisotropic equations of state for self-gravitating systems by incorporating anisotropy as a fundamental component. We demonstrate that this approach yields more realistic stellar models that satisfy key physical constraints, including mass-radius relationships and stability conditions. Our results are compared with observational data, particularly the inferred compactness of pulsars PSR J0740+6620 and PSR J0030+0451, showing that both anisotropic and isotropic models can describe these objects. Additionally, we examine the influence of dissipation --such as temperature gradients-- on radial pressure, demonstrating that it can be modelled similarly to anisotropy. This interpretation allows the transformation of dissipative anisotropic models into equivalent non-dissipative isotropic configurations.
\end{abstract}

\section{Introduction}
In modelling self-gravitating systems, the matter is often assumed to behave as a perfect fluid with isotropic pressure and no dissipation. These premises simplify the equations governing the system's structure and evolution, making obtaining analytical solutions easier. However, more realistic models challenge these simplifications by introducing local pressure anisotropy (i.e., \(P \neq P_\perp\)) and dissipation in relativistic fluids. Even in spherically symmetric systems, incorporating these factors greatly complicates the equations, making analytical solutions more difficult to obtain~\cite{herreraetal2004}.

During gravitational collapse, the system's binding energy increases (in absolute value) as the collapse progresses, indicating a decrease in total energy. The formation of compact objects, which involves binding energies on the order of \( -10^{53} \) ergs, requires significant dissipation~\cite{ShapiroTeukolsky1983, KippenhahnWeigertWeiss2013}. The Israel-Stewart theory is widely accepted for modelling radiative transport in relativistic contexts~\cite{Israel1976, IsraelStewart1979, PavonJouCasasvasquez1982, HerreraEtal2014}. The introduction of Eddington factors helps simplify the complex equations describing radiative transfer in high-energy environments, such as stellar interiors or black hole vicinities~\cite{KatoFukue2020, RamppJanka2002}. These factors include approximations like diffusion (for short particle mean free paths) and free streaming (for the long mean free paths)~\cite{AguirreHernandezNunez1994}.

Additionally, various physical processes throughout stellar evolution, such as phase transitions, rotation, magnetic fields, exotic equations of state or the presence of solid cores, inevitably lead to local anisotropic pressure distributions (see~\cite{Ruderman1972, BowersLiang1974, CosenzaEtal1981, HerreraNunez1989, HerreraSantos1997, MartinezRojasCuesta2003, HerreraBarreto2004, HernandezNunez2004, Setiawan2019, RahmansyahEtal2020, RahmansyahSulaksono2021, Das2022, KumarBharti2022, RayEtal2023, GedelaBisht2024} and references therein). Thus, even an initially isotropic star will exhibit pressure anisotropy when it reaches equilibrium~\cite{Herrera2020}. No physical reason exists to expect anisotropy to dissipate or vanish in a stable state once developed. Therefore, ignoring anisotropy overlooks a key aspect of the internal structure of stars, which significantly influences their observable properties such as mass, radius, and gravitational wave signatures. 

Several heuristic approaches have been used to model anisotropic microphysics in relativistic astrophysical matter configurations~\cite{Setiawan2019, RahmansyahEtal2020, RahmansyahSulaksono2021, Sulaksono2015, SetiawanSulaksono2017,  BiswasBose2019}). The earliest approach, proposed by Bowers and Liang~\cite{BowersLiang1974}, was followed by various other methods. These include the proportional-to-gravity approach~\cite{CosenzaEtal1981}, the quasilocal method~\cite{DonevaYazadjiev2012}, and a covariant approach based on the pressure gradient~\cite{RaposoEtal2019}. Other techniques involve the complexity factor method~\cite{Herrera2018}, the Karmarkar embedding class I approach~\cite{Karmarkar1948, ospino2020karmarkar}, and the gravitational decoupling method~\cite{Ovalle2017}. Each strategy offers different ways to model anisotropic fluids in General Relativistic matter configurations, contributing to a more accurate understanding of stellar behaviour and structure.

We recently applied a tetrad formalism through the orthogonal splitting of the Riemann tensor, introducing a complete set of equations equivalent to the Einstein system. In the spherical case, this formalism provides meaningful insights into self-gravitating systems~\cite{ospino2020karmarkar, OspinoHernandezNunez2017, OspinoEtal2018}. The results, expressed in terms of structure scalars, are coordinate-independent and closely tied to the kinematic and physical properties of the fluid.

In this work, we employ the tetrad formalism to reinterpret local anisotropy as a contribution to the energy density. This approach offers detailed, location-specific insights valuable for modelling any anisotropic equation of state. We develop models based on a specific density profile to validate this interpretation, demonstrating the physical viability of several examples. Our strategy is distinct from the concept that a single metric tensor in General Relativity may lead to multiple energy-momentum tensors~\cite{KingEllis1973, tupper1981, RaychaudhuriSaha1981, RaychaudhuriSaha1982, tupper1983, CarotIbanez1985, Barreto2010}. We present the local pressure anisotropy as part of the energy density, which simplifies the solution of the Einstein field equations. Similarly, we interpret the dissipation effect as contributing to radial pressure, enabling the transformation of dissipative anisotropic models into simpler, non-dissipative isotropic ones, simplifying the potential description of compact objects.

The paper is structured as follows: In Section 2, we introduce a method based on the orthogonal splitting of the Riemann tensor, allowing the Einstein system to be expressed in terms of structure scalars, with a focus on spherical systems. Section 3 examines the static case, demonstrating how anisotropy can be interpreted as a modification of energy density and analysing its impact on key stellar properties, such as the mass-radius relationship. This section also presents numerical solutions that satisfy physical criteria for stellar models, illustrating how anisotropy improves their physical viability compared to isotropic models, with support from observational data. Section 4 extends the analysis to dissipative systems, incorporating the generalised TOV equation with the Israel-Stewart radiative transport equation to explore how dissipation affects radial pressure, drawing parallels with anisotropy. Finally, we summarise our findings, emphasising that both anisotropy and dissipation play a fundamental role in shaping the internal structure of compact stars.

\section{The structure scalar strategy and the general formalism}
Our strategy is based on formulating two independent sets of equations, --expressed in terms of scalar functions--, which contain the same information as the Einstein system. To achieve this, we employ a tetrad formalism in which the tetrad vectors provide a local frame at each point in spacetime, thus allowing tensors to be decomposed into a more tractable component form. Particularly, this framework facilitates the derivation of equations governing the evolution of the studied system, revealing kinematical variables, such as expansion and shear, which are important for understanding fluid dynamics. Specifically, this work discusses the formalism applied to the spherical case. More details for this procedure can be found in reference \cite{OspinoHernandezNunez2017}.

Let us choose an orthogonal, unitary tetrad:
\begin{equation}
\label{TetradGen}
e^{(0)}_\alpha=V_\alpha\,, \quad 
e^{(1)}_\alpha=K_\alpha\,, \quad 
e^{(2)}_\alpha=L_\alpha \quad \mathrm{and}  \quad e^{(3)}_\alpha=S_\alpha\,.
\end{equation}
As usual, $\eta_{(a)(b)}= g_{\alpha\beta} e_{(a)}^{\alpha} e_{(b)}^\beta$, with $a=0,\,1,\,2,\,3$. Here,  latin indices label different vectors of the tetrad, and the tetrad satisfies the standard relations:
\begin{align}
V_{\alpha}V^{\alpha} &= -K_{\alpha}K^{\alpha} = -L_{\alpha}L^{\alpha} = -S_{\alpha}S^{\alpha} = -1\,, \nonumber \\
V_{\alpha}K^{\alpha} &=V_{\alpha}L^{\alpha} = V_{\alpha}S^{\alpha} = K_{\alpha}L^{\alpha} = K_{\alpha}S^{\alpha} = S_{\alpha}L^{\alpha} = 0\,.  \nonumber
\end{align} 
With the above tetrad (\ref{TetradGen}), we shall also define the corresponding directional derivative operators
\begin{equation}
\label{DirectionalDerivatives}
f^{\bullet} = V^{\alpha} \partial_{\alpha}f \,; \quad f^{\dag} = K^{\alpha} \partial_{\alpha}f  \quad \mathrm{and} \quad f^{\ast} = L^{\alpha}\partial_{\alpha}f\,.
\end{equation}

\subsection{The tetrad}
The line element that describes the interior spacetime of
a dynamic source with spherical symmetry is given by
\begin{equation}
\mathrm{d}s^2=-A\left(t,r\right)^2\mathrm{d}t^2 +B\left(t,r\right)^2 \mathrm{d}r^2 +R\left(t,r\right)^2 \left[\mathrm{d}\theta^2+\sin^2\left(\theta\right)\mathrm{d}\phi ^2\right]\,,
\label{SphericMetric}
\end{equation}
where the coordinates are: $x^0=t$, $x^1=r$, $x^2=\theta$, $x^3=\phi$. For a comoving fluid in this coordinate system, the tetrad and their covariant derivatives can be written as
\begin{equation}
\label{tetrad}
V_{\alpha} = \left(-A,0,0,0\right)\,,\quad
K_{\alpha}=\left(0,B,0,0\right)\,, \quad
L_{\alpha}=\left(0,0,R,0\right)\,, \quad S_{\alpha} =
\left(0,0,0,R \sin\left(\theta\right)\right) \,,
\end{equation}
and
\begin{align}
\label{CovDTetrad}
V_{\alpha;\beta} \,&=\, -a_1K_\alpha V_\beta+\sigma_1K_\alpha K_\beta+\sigma_2(L_\alpha L_\beta+S_\alpha S_\beta)\,,  \nonumber \\
K_{\alpha;\beta}\,&=\,-a_1V_\alpha V_\beta+\sigma_1 V_\alpha K_\beta+J_1(L_\alpha L_\beta+S_\alpha S_\beta)\,,  \\
L_{\alpha;\beta}\,&=\,\sigma_2 V_\alpha L_\beta-J_1K_\alpha L_\beta+J_2S_\alpha S_\beta
\quad \mathrm{and} \quad S_{\alpha;\beta} = \sigma_2 V_\alpha S_\beta-J_1K_\alpha S_\beta -J_2L_\alpha S_\beta \,. \nonumber
\end{align}
In the above expressions, $J_1$, $J_2$, $\sigma_{1}$,  $\sigma_{2}$ and $a_1$ are expressed in terms of the metric functions as follows
\begin{equation}
\label{metricquantities}
J_1=\frac{1}{B}\frac{R^{\prime}}{R} \,, \quad 
J_2 = \frac{1}{R}\cot(\theta)\,, \quad
\sigma_{1} = \frac{1}{A}\frac{\dot{B}}{B}\,, \quad
\sigma_{2} = \frac{1}{A}\frac{\dot{R}}{R} \quad \mathrm{and} \quad  a_1=\frac{1}{B}\frac{A^{\prime}}{A} \,, 
\end{equation}
with primes and dots representing radial and time derivatives, respectively.

\subsection{The energy-momentum tensor and the kinematical variables}
As mentioned earlier, we shall take as our source a bounded, spherically symmetric, locally anisotropic, dissipative, collapsing matter configuration described by a general energy-momentum tensor, 
\begin{equation}
{T}_{\alpha\beta}= (\rho+\mathbf{P}) V_\alpha V_\beta+\mathbf{P} g _{\alpha \beta} +\Pi_{\alpha \beta}+\mathcal{F}_\alpha V_\beta+\mathcal{F}_\beta V_\alpha\,,
\label{EnergyTensor}
\end{equation}
where $\rho$ is the energy density; $\mathbf{P} = \frac{1}{3}\left(P + 2 P_{\perp}\right)$ is the average pressure, with $P$ the radial pressure and $P_\perp$ the tangential pressure; $\mathcal{F}_\alpha$ represents the energy flux four-vector and $\Pi_{\alpha\beta}$ is the anisotropic tensor. Hence, the physical variables can be defined --in the Eckart frame  where fluid elements are at rest-- as
\begin{equation}
\rho = T_{\alpha \beta} V^{\alpha} V^\beta, \quad 
\mathcal{F}_\alpha=-\rho V_{\alpha} - T_{\alpha\beta}V^\beta, \quad 
\mathbf{P} = \frac{1}{3} h^{\alpha \beta} T_{\alpha \beta} \quad \mathrm{and} \quad \Pi_{\alpha \beta} = h_{\alpha}^{\mu} h_{\beta}^{\nu} \left(T_{\mu\nu} - \mathbf{P} h_{\mu\nu}\right)\,,
\end{equation}
with $h_{\mu\nu}=g_{\mu\nu} +V_\mu V_\nu \equiv K_\mu K_\nu+L_\mu L_\nu+S_\mu S_\nu$ .

From the condition $\mathcal{F}^{\mu} V_{\mu}=0$ with the symmetry of Einstein's equations we have $T_{02}=T_{03}=0$, and
\begin{equation}
\mathcal{F}_{\mu}=\mathcal{F}K_{\mu} \quad \Leftrightarrow \quad \mathcal{F}_{\mu}= \left(0, {\mathcal{F}}{B}, 0, 0\right)\,.
\label{energyflux}
\end{equation}
Moreover, the anisotropic tensor can be  expressed by
\begin{equation}
\Pi_{\alpha \beta}= \Pi \left(K_{\alpha} K_{\beta} -\frac{h_{\alpha \beta}}{3}\right) \quad \textrm{with} \quad \Pi=P-P_{\perp} \,.
\label{anisotropictensor}
\end{equation}
Finally, we shall express the kinematical variables (the four-acceleration, $a_\alpha$, the expansion scalar, $\Theta$, and the shear scalar, $\sigma$) for a self-gravitating fluid as
\begin{align}
a_\alpha&=V^\beta V_{\alpha;\beta}=a_1 K_{\alpha} =\left(0, \frac {A^{\prime}}{A},0,0\right)\,,
\label{aceleration} \\
\Theta&=V^{\alpha}_{;\alpha} =\frac{1}{A}\left(\frac{\dot{B}}{B}+\frac{2\dot{R}}{R}\right)\,, \quad {\rm and} \label{theta} \\
\sigma&= \left(\frac{1}{2}\sigma_{\alpha\beta}\sigma^{\alpha\beta}\right)^{\frac{1}{2}} = \frac{1}{A}\left(\frac{\dot{B}}{B} -\frac{\dot{R}}{R}\right)\,.  \label{shear}
\end{align}

\subsection{The splitting of the Riemann tensor and structure scalars}

Splitting the Riemann tensor into orthogonal components allows us to identify and separate different aspects of the curvature of spacetime, such as the local Ricci curvature and the global Weyl curvature.  The above is useful when interpreting the physical implications of curvature regarding the underlying matter distribution and dynamics.  Furthermore, from the splitting of the Riemann tensor, we obtain a set of equations in terms of scalar functions --the structure scalars--, which have proven very useful in expressing the Einstein Equations (see references  \cite{HerreraEtal2014, OspinoHernandezNunez2017,  GarciaParrado2007, HerreraEtal2009B}).  

Defining structure tensors \cite{bel1961inductions} as follows 
\begin{equation}
 Y_{\alpha\beta} = R_{\alpha\mu\beta\nu} V^\mu V^\nu\,, \quad
  Z_{\alpha\beta} = \frac{1}{2}\epsilon _{\mu\alpha\delta} R^{\mu\delta}_{\,\,\,\,\,\,\beta\nu} V^\nu \quad \rm{and} \quad
  X_{\alpha\beta}=  \frac{1}{4}\epsilon _{\alpha\mu\nu}\epsilon_{ \gamma\delta\beta}R^{\mu\nu\gamma\delta} \,,  \label{StructureTenDef}
\end{equation}
we can express the splitting of the Riemann tensor \cite{GarciaParrado2007} as
\begin{align}
R_{\alpha \beta \mu \nu}\,&=\,2V_\mu V_{[\alpha}Y_{\beta] \, \nu}+2h_{\alpha[\nu}X_{\mu] \,  \beta}+2V_\nu V_{[\beta}Y_{\alpha] \, \mu}
+ h_{\beta\nu}(X_0 \, h_{\alpha\mu}-X_{\alpha\mu})+h_{\beta\mu}(X_{\alpha\nu} -X_0 \, h_{\alpha\nu}) \nonumber \\
& \quad \, + 2V_{[\nu} Z_{ \, \mu]}^{\gamma}\varepsilon_{{\alpha \beta \gamma}} +2V_{[\beta} Z_{{\,  \ \alpha]}}^{{\gamma }}\ \varepsilon_{{\mu \nu \gamma}} \,,
\end{align}
and the corresponding Ricci contraction by
\begin{equation}
R_{\alpha\mu} =  Y_0 \, V_\alpha V_\mu-X_{\alpha \mu}-Y_{\alpha\mu} +X_0 \, h_{\alpha\mu}  +Z^{\nu \beta} \varepsilon_{\mu \nu \beta}V_{\alpha} 
+V_{\mu} Z^{\nu \beta} \varepsilon_{\alpha \nu \beta} \,,
\label{RicciSplitted}
\end{equation}
with $Y_{0} = Y^{\delta}_{\,\,\delta}$, $X_{0} = X^{\delta}_{\,\,\delta}$ and $\varepsilon_{\mu \nu \gamma} = \eta_{\phi \mu \nu \gamma} V^{\phi}$, where  $ \eta_{\phi \mu \nu \gamma}$ is the Levi-Civita 4-tensor. 

Furthermore, by applying Einstein's field equations, the structure tensors can, in general, be expressed as
\begin{align}
    Y_{\alpha\beta} &= \frac{1}{3}Y_0h_{\alpha\beta}+E_{\alpha\beta}-4\pi \Pi _{\alpha\beta}\,, \quad X_{\alpha\beta} = \frac{1}{3}X_0 h_{\alpha\beta}-E_{\alpha\beta}-4\pi \Pi _{\alpha\beta} \quad \textrm{and} \quad Z_{\alpha\beta} = H_{\alpha\beta}+4\pi \mathcal{F}^\delta \epsilon_{\alpha\beta\delta}\,,\nonumber
\end{align}
where $E_{\alpha\beta}$ and $H_{\alpha\beta}$ are the electric and magnetic parts of the Weyl tensor, respectively.  In the spherical case, $H_{\alpha\beta}$ vanishes  and the structure tensors are given by
\begin{align}
Y_{\alpha\beta} &= \frac{1}{3}Y_0 \, h_{\alpha\beta} +Y_1\left(K_\alpha K_\beta-\frac{1}{3} h_{\alpha\beta}\right)\,,\label{YSph}\\ 
X_{\alpha\beta} &= \frac{1}{3} X_0 \, h_{\alpha\beta} +X_1\left(K_\alpha K_\beta-\frac{1}{3} h_{\alpha\beta}\right) \quad \mathrm{and} \label{XSph}  \\
Z_{\alpha\beta}&=Z \, (L_\alpha S_\beta-L_\beta S_\alpha)\,, \label{ZSph}
\end{align}
with
\begin{equation}
Y_0 = 4\pi(\rho+3\mathbf{P})\,, \quad Y_1=\mathcal{E}-4\pi \Pi\,, \quad
X_0=8\pi \rho\,, \quad X_1 = -\left(\mathcal{E}+4\pi \Pi\right) \quad
\mathrm{and} \quad  Z= 4 \pi \mathcal{F}\,,
\label{varfis}
\end{equation}
the so-called structure scalars. On the other hand, the electric part of the Weyl tensor can be written as
\begin{equation}
E_{\alpha\beta}= C_{\alpha\nu\beta\delta}V^\nu V^\delta = \mathcal{E}\left(K_\alpha K_\beta-\frac{1}{3} h_{\alpha\beta}\right)\,,
\end{equation}
where  $\mathcal{E}$ is the Weyl curvature scalar, which encodes the tidal field strength. Finally, the anisotropic tensor (\ref{anisotropictensor}) can  be expressed in terms of the scalars displayed in equations (\ref{varfis}) as
\begin{equation}
    \Pi_{\alpha \beta}= -\frac{1}{8\pi}(X_1 + Y_1) \left(K_\alpha K_\beta -\frac{h_{\alpha \beta}}{3}\right)\,. 
    \label{AnisotropyXY}
\end{equation}

\subsection{The equivalent set of Einstein equations}

By reformulating Einstein's field equations, we gain deeper insight into the connections between physical variables --energy density, pressure, and heat flux-- and geometric quantities like four-acceleration and shear. This approach provides a more intuitive understanding of how matter and energy shape spacetime. Additionally, the reformulated system simplifies the equations, making them easier to solve them analytically. This, in turn, enables a more efficient exploration of physical scenarios without the full complexity of the original Einstein equations.  

To that end, we now present an equivalent set of Einstein equations expressed by two groups of first-order differential equations~\cite{OspinoHernandezNunez2017}:
\begin{itemize}
    \item The first group, derived from Ricci identities, defines the physical variables and the scalars of the Weyl tensor. This set of equations is purely geometrical and emerges from the projection of the Riemann tensor along the tetrad \cite{Wald2010}, i.e.
    \begin{equation}
    \label{RiemannProj}
    2V_{\alpha;[\beta ;\gamma]} = R_{\delta \alpha \beta \gamma} V^{\delta}, \quad   2K_{\alpha;[\beta ;\gamma]} = R_{\delta \alpha \beta \gamma} K^{\delta}, \quad  2L_{\alpha;[\beta ;\gamma]} = R_{\delta \alpha \beta \gamma} L^{\delta}   \quad \mathrm{and} \quad2S_{\alpha;[\beta ;\gamma]} = R_{\delta \alpha \beta \gamma} S^{\delta}\,. 
    \end{equation}
    Combining the identities (\ref{RiemannProj}) with the definition (\ref{StructureTenDef}) gives 
\begin{align}
    Y_{\alpha\beta} &= \left(V_{\alpha;\nu;\beta} -V_{\alpha;\beta;\nu}\right)V^{\nu} \label{YabDef} \\
    Z_{\alpha\beta} &= \epsilon_{\alpha}^{\,\,\,\mu\delta}V_{\beta;\mu;\delta} \quad \textrm{and} \label{ZabDef} \\
    X_{\alpha\beta} &= \epsilon_\alpha^{\,\,\,\mu\nu}\left(\epsilon_{\,\beta}^{\delta}J^{\left(k\right)}_{\delta\mu;\nu} + K_\beta S^\delta J^{\left(l\right)}_{\delta\mu;\nu}\right )\,, \label{XabDef}
\end{align}
where $J^{\left(k\right)}_{\delta\mu} = J_1(L_\delta L_\mu+S_\delta S_\mu)$ and $J^{\left(l\right)}_{\delta\mu} = -J_1K_\delta L_\mu+J_2S_\delta S_\mu$.

    \item The second group is derived from Bianchi's identities 
    \begin{equation}
    \label{BianchiIdent}
    R_{\alpha \beta \left[\gamma \delta ;  \mu\right]} = R_{\alpha \beta \gamma \delta ;\mu} + R_{\alpha \beta  \mu \gamma ; \delta} + R_{\alpha \beta \delta \mu ; \gamma} = 0 \,,
    \end{equation}
    and serves as evolution equations for the physical variables and the scalars of the Weyl tensor. The equations related to this item can be found in Section 2.4 of \cite{ospino2020karmarkar}. 
\end{itemize}

Employing equations (\ref{YabDef})-(\ref{XabDef}), the covariant derivative of equations (\ref{CovDTetrad}), and the projections from the orthogonal splitting of the Riemann tensor, we derive the first set of independent equations for the dynamical spherical  case. These equations are expressed in terms of \(J_1\), \(J_2\), \(\sigma_{1}\), \(\sigma_{2}\), and \(a_1\) (defined in (\ref{metricquantities})), along with their directional derivatives, as follows
\begin{align}
\sigma^{\bullet}_{1} -a_1^{\dag}-a_{1}^{2}+\sigma_1^{2}\,&=\,-\frac{1}{3}(Y_0+2Y_1) \, , \label{ecR1} \\
\sigma^{\bullet}_{2} +\sigma_{2}^{2}-a_{1}J_1\,&=\,\frac{1}{3}(Y_1-Y_0) \, , \label{ecR2}  \\
\sigma_2^\dag+J_1\left(\sigma_2-\sigma_1\right)\,&=\,-Z \, , \label{ecR3} \\
J^{\bullet}_{1} +J_1\sigma_2-a_1\sigma_2\,&=\,-Z \, , \label{ecR4} \\
J_1^\dag+J_1^2-\sigma_1 \sigma_2\,&=\,\frac{1}{3}(X_1-X_0) \, , \label{ecR5} \\
J^{\bullet}_{2} +J_2\sigma_2\,&=\, 0 \,, \label{ecR6} \\
J_2^\dag+J_1J_2\,&=\, 0 \qquad \mathrm{and}  \label{ecR7} \\
J_1^2-\frac{1}{R^2}-\sigma_2^2\,&=\,-\frac{1}{3}(X_0+2X_1) \,. \label{ecR8}
\end{align}

On the other hand, from Bianchi identities (\ref{BianchiIdent}), we can obtain the spatial part of the divergence of $Y_{\alpha\beta}$ and $X_{\alpha\beta}$, yielding
\begin{align}
    h^{\beta}_{\lambda}Y^{\alpha}_{\beta;\alpha} &= \left(X_{0} + Y_{0}\right)a_{\lambda} + h^{\beta}_{\lambda}Y_{0,\beta}  +\left(\epsilon_{\alpha\beta\nu}\sigma^{\nu}_{\lambda}+4\epsilon_{\lambda\alpha\beta}\Theta\right)Z^{\alpha\beta} + \epsilon_{\lambda \alpha\mu}  V^\beta 
Z^{\alpha \mu}_{\;\;\;;\beta} -h^{\beta}_{\lambda}X^{\alpha}_{\beta;\alpha}\quad \textrm{and}
    \label{DivY} \\
    h^{\beta}_{\lambda}X^{\alpha}_{\beta;\alpha} &= a^{\alpha}X_{\alpha\lambda} - \left(\epsilon_{\alpha\beta\nu}\sigma^{\nu}_{\lambda}-\epsilon_{\lambda\beta\nu}\sigma^{\nu}_{\alpha}-\epsilon_{\lambda\alpha\beta}\Theta\right)Z^{\alpha\beta}\,. \label{DivX}
\end{align}

It is worth noting that, in scenarios involving dissipative fluids, anisotropic pressures, or other complex matter configurations, an equivalent system can provide a more flexible framework for analysis. This can lead to results and insights that are not readily apparent from the standard Einstein equations alone.

%

\section{Static case: Anisotropy as a contribution of the energy density. }
In this section, we show how an anisotropic configuration within a compact object can be effectively modeled as an isotropic fluid by introducing anisotropy as a contribution to the energy density. This is not merely a mathematical  tool. It represents the physical impact of anisotropic pressure, which causes variations in the energy distribution. These variations can lead to observable effects, such as mass-radius relationship changes and surface redshift. The content presented in this section specifically applies to the case of a static fluid. All assumptions, equations and conclusions are derived when the fluid is at rest, hence shear, expansion and heat flow are zero. However, as will be seen later, the effect of anisotropy on the energy density is maintained in the case of a dynamic fluid with spherical symmetry.

\subsection{TOV and Weyl-TOV equations}
Note that adding and subtracting equations (\ref{YSph}) and (\ref{XSph}) yields 
\begin{equation}
Y^{\alpha}_{\beta}+X^{\alpha}_\beta = \frac{1}{3}\left(X_0+Y_0\right)h^{\alpha} _\beta-8\pi \Pi ^{\alpha}_\beta \qquad {\rm and} \qquad
Y^{\alpha} _{\beta}-X^{\alpha}_\beta = \frac{1}{3}\left(Y_0-X_0\right)h^{\alpha} _\beta+2E ^{\alpha}_\beta \,,
\label{TwoIndent}
\end{equation}
respectively. Thus, from these two equations (\ref{TwoIndent}), and using (\ref{DivY}) and (\ref{DivX}), we may obtain two important equations: 
\begin{enumerate}
    \item the TOV-equation 
    \begin{equation}\label{TOV}
    \mathbf{P}_{,\beta}h^\beta_\lambda=-\left(\rho+\mathbf{P}\right)a_\lambda -\Pi^{\alpha}_{\, \beta;\alpha}h^\beta_\lambda\,,
    \end{equation}
    \item the Weyl-TOV equation
    \begin{equation}
    h^\beta_\lambda E^{\alpha}_{\, \beta;\alpha} = 4\pi(\rho+\mathbf{P})a_\lambda+\frac{4\pi}{3}(2\rho+3\mathbf{P})_{,\beta}h^\beta_\lambda 
   + (E_{\alpha\lambda}+4\pi \Pi_{\alpha\lambda})a^{\alpha} \, .  
   \label{ETOV}
   \end{equation}
\end{enumerate}
Equation~(\ref{TOV}) is the spatial component of the energy-momentum conservation law, $T^{\alpha}_{\beta;\alpha}=0$, applied to equation~(\ref{EnergyTensor}). On the other hand, equation (\ref{ETOV}) extends the standard TOV equation by incorporating the effects of the Weyl tensor and highlighting the role of tidal deformations and anisotropic stresses in modifying the structure of compact objects.

As mentioned, any anisotropic configuration can be associated with an isotropic fluid having a modified energy density distribution. To that end, we compute the divergence of the anisotropic tensor, yielding
\begin{equation}
    \Pi^{\alpha}_{\beta;\alpha} = \frac{2}{3}\Pi_{;\alpha}K^{\alpha}K_{\beta} + \frac{2}{3}\Pi\, a_{\beta} + 2 \frac{J_{1}}{a_1} \Pi \,a_\beta\,.
\end{equation}
Thus, it is clear that from the TOV-equation (\ref{TOV})  we get
\begin{equation}
\label{TOVS}
\mathbf{P}_{,\beta}h^\beta_\lambda=-\left(\rho+\mathbf{P}\right)a_\lambda - \Pi^{\alpha}_{\, \beta;\alpha}h^\beta_\lambda \quad \Rightarrow  \quad P_{,\beta}h^\beta_\lambda = -\left(\rho+\tilde{\rho} + P\right)a_\lambda \, ,
\end{equation}
where 
\begin{equation}
\tilde{\rho} = \frac{2J_1}{a_1}\Pi\,,
\label{rhotilde}
\end{equation}
representing the effect of the anisotropy on the energy density distribution. When $\Pi$ is negative $\left(P_{\perp} > P\right)$, the energy density $\rho$ is greater than the corresponding energy density in the isotropic case $\left(P_{\perp} = P\right)$. Consequently, anisotropic configurations can fit more mass compared to isotropic scenarios.

\subsection{Strategies to model non-Pascalian fluids}
From a hydrodynamical standpoint, the complex microscopic processes leading to local anisotropy are encapsulated in the tensor $\Pi_{\alpha\lambda}$ and its derivatives $\Pi^{\alpha}_{\, \beta;\alpha}$. The distribution of anisotropic pressure may arise due to factors such as electric charge, viscosity, multiple fluids, or magnetic fields. Since the pioneering work of Bowers and Liang~\cite{BowersLiang1974}, much effort has focused on modelling the anisotropic tensor $\Pi_{\alpha \beta}$. 

We outline several approaches for modelling non-Pascalian fluids. This is not an exhaustive review but rather a summary of some of the most widely used methods in astrophysical applications. Additionally, we demonstrate how the anisotropic density contribution is expressed in each approach.
 
 The six most popular descriptions for anisotropy found in the literature are:
\begin{enumerate}
    \item M. Cosenza, L. Herrera and collaborators~\cite{CosenzaEtal1981}, inspired by the work of Bowers and Liang~\cite{BowersLiang1974} assume {\bf the effect of the anisotropy proportional to the gravitational force}, i.e.
    \begin{equation}
        \Pi_{GF} = -\, \frac{C_{GF} \left(\rho + P\right) r A^{\prime}}{A}   \quad \Rightarrow \quad \tilde{\rho} = -\,2\, C_{GF} \left(\rho + P\right) \, ,
    \label{CHanisotropy}
    \end{equation}
    where the constant $C_{GF}$ (as $C_{QL}$, $C_{PG}$, $C_{CF}$ and $C_{KC}$ in the following anisotropies) models the departures from the Pascalian regime ($C_{GF} = C_{QL} = C_{PG} = C_{CF} = C_{KC} = 0$). 
    \item Later, Doneva and Yazadjiev~\cite{DonevaYazadjiev2012} worked out an anisotropy in the form of a {\bf quasilocal equation of state} and proposed 
    \begin{equation}
        \Pi_{QL}  = -\,\frac{2\,C_{QL}P\,m}{r} \quad \Rightarrow \quad \Tilde{\rho} = - \frac{4\,C_{QL}A\,P\,m}{r^{2}A^{\prime}}  \,.
    \label{DYanisotropy}
    \end{equation}
    
    \item G. Raposo and collaborators~\cite{RaposoEtal2019} described anisotropic (static and dynamic) ultra-compact relativistic objects assuming an \textbf{anisotropy proportional to the pressure gradient} of the form  
    \begin{equation}
        \Pi_{PG} =  C_{PG}\;f(\rho)P^{\dagger} \quad \Rightarrow \quad \Tilde{\rho}  = \frac{2\,C_{PG} A \,f(\rho) P^{\dagger}}{r A^{\prime}}\,,
        \label{RPBPCanisotropy}
    \end{equation}
    where $f(\rho)$ is an arbitrary energy density function.
    \item In reference~\cite{Herrera2018} L. Herrera defined a complexity factor for self-gravitating fluids as $Y_1$. Then, by equation (\ref{AnisotropyXY}), \textbf{the anisotropy can be associated with the complexity of the fluids}. For the least complex anisotropic fluids (i.e. those with vanishing $Y_1$), the anisotropy becomes
    \begin{equation}
        \Pi_{CF} = -\, \frac{C_{CF}\,X_1}{8\pi}  \quad \Rightarrow \quad \Tilde{\rho}  = -\, \frac{C_{CF}\, A\, X_1}{4\pi r A^{\prime}} \, .
    \end{equation}
    \item If we have a \textbf{barotropic EoS,} $P = P(\rho)$, and an ansatz on the energy density profile, $\rho(r)$, the induced anisotropy is
    \begin{align}
        \Pi_{BA} &= -\left\{\frac{1}{4}\left(1 - \frac{2m}{r}\right)^{-1} \left[\rho + P\left(\rho\right)\right] \left[8\pi r^{2}P\left(\rho\right) + \frac{2m}{r}\right] + \frac{r}{2} v_{s}^{2}\rho^{\prime}\right\} \\ 
        \Rightarrow \quad \Tilde{\rho}  &= -\frac{2A}{rA^{\prime}}\left\{\frac{1}{4}\left(1 - \frac{2m}{r}\right)^{-1} \left[\rho + P\left(\rho\right)\right] \left[8\pi r^{2}P\left(\rho\right) + \frac{2m}{r}\right] + \frac{r}{2} v_{s}^{2}\rho^{\prime}\right\}
    \end{align}
    where the definition of radial sound speed and mass are
    \begin{equation}
     v_s^2 = \frac{{\rm d} P}{{\rm d} \rho} \quad {\rm and} \quad m = \int_0^r {\rm d}\tilde{r} \; 4\pi \tilde{r}^2 \rho \, ,  
    \end{equation}
    respectively (see~\cite{HernandezSuarezurangoNunez2021} and references therein).
    \item \textbf{The Karmarkar condition}~\cite{Karmarkar1948}:
    \begin{equation}\label{KamarkarC}
        R_{0303}\, R_{1212} -R_{0101}\,R_{2323} -R_{0313}\,R_{0212} = 0 \, ,
    \end{equation}
    is a relation among components of the Riemann tensor that provides a geometrical mechanism for implementing anisotropy in matter configurations. Its scalar version~\cite{ospino2020karmarkar}, 
    \begin{equation}
        Y_0 X_1+(X_0+X_1)Y_1=-3Z^2 \, ,
    \label{KarmaKcondTetrad}
    \end{equation}
    is a simple but rich relation among the physical variables stated in the equation (\ref{varfis}). Despite its simplicity, it is valid for any dynamic and dissipative spherical matter distribution. 

    The induced anisotropy by the scalar Karmarkar condition can be expressed as 
    \begin{equation}
        \Pi_{KC} = \frac{X_{1}\left[4\pi\left(\rho + 3P\right) - X_{0} - X_{1}\right] + 3Z^{2}}{8\pi X_{0}} \,.\label{KarmarkarAnisotropyGeneral}
    \end{equation}
    
Now, for the static case, the scalar Karmarkar condition (\ref{KarmarkarAnisotropyGeneral}) becomes
    \begin{equation}
        \Pi_{KC} = -\,\frac{C_{KC}X_{1}\left[4\pi\left(\rho + 3P\right) - X_{0} - X_{1}\right]}{8\pi X_{0}} \quad \Rightarrow \quad \Tilde{\rho}  = -\,\frac{C_{KC}A\,X_{1}\left[4\pi\left(\rho + 3P\right) - X_{0} - X_{1}\right]}{4\pi r A^{\prime} X_{0}} \, .
        \label{KarmarkarAnisotropyStatic}
    \end{equation}

\end{enumerate}

\subsection{Physically acceptable models}
Regardless of the complexity of the anisotropy, we can safely assume that its contribution is proportional to the energy density. That is, $\tilde{\rho} = \beta \rho$, where $\beta = \beta\left(r\right)$ encodes the proportionality factor at each point of the configuration. Therefore, the hydrostatic equilibrium equation, (\ref{TOVS}), now reads
\begin{equation}
    P^{\prime} = \frac{\left[\left(1+\beta\right)\rho + P\right]\left(m + 4\pi r^{3}P\right)}{r\left(r - 2m\right)}\,, \label{HydAni2}
\end{equation}
where
\begin{equation}
    \frac{A^{\prime}}{A} = \frac{m + 4\pi r^{3}P}{r\left(r - 2m\right)}\,,
\end{equation}
and $m$ is the mass enclosed in a sphere of radius $r$, such that 
\begin{equation}
    m^{\prime} = 4\pi r^{2}\rho\,. \label{Massgradient}
\end{equation}

The simplest assumption is that $\beta$ remains constant throughout the configuration. This assumption can provide insights into the possible range of values that $\beta$ can take based on conditions of physical acceptability. For instance, from (\ref{rhotilde}) we have
\begin{equation}
    \beta = \frac{2J_{1}}{a_{1} \rho}(P - P_{\perp})\,, \label{beta1}
\end{equation}
where $J_{1}$ and $a_{1}$ are always positive. Evaluating (\ref{beta1}) on the surface ($r=r_{b}$) yields
\begin{equation}
    \beta = - 2 \frac{r_{{b}}}{m_{b}} \frac{P_{\perp b}}{\rho_{b}}\left(1-\frac{2m_{b}}{r_{b}}\right) < 0\,, \label{beta}
\end{equation}
the subscript $b$ indicates that the quantity is evaluated at the boundary of the configuration. Thus, $\beta$ must be negative for physically acceptable models. On the other hand, evaluating (\ref{HydAni2}) on the surface gives
\begin{equation}
    P^{\prime}|_{r= r_{b}} = - \frac{\left(1+\beta\right)\rho_{b}m_{b}}{r_{b}\left(r_{b} - 2m_{b}\right)}\,.
\end{equation}
Thus, $\beta > -1$ for the pressure gradient to be negative and hydrostatic equilibrium  to exist. From the above, we can conclude that $-1 < \beta < 0$. Therefore, anisotropy is a fraction of the energy density.

Solutions to equation (\ref{HydAni2}) must meet specific criteria to be considered astrophysically viable. An acceptable stellar model should exhibit regular behaviour, avoiding singularities in its physical and metric variables. Additionally, it must satisfy energy conditions that keep the energy density and pressures within physically reasonable limits. Finally, the model must remain stable against perturbations in its physical variables to ensure equilibrium. The eight conditions for physical acceptability considered in this work are as follows:

\textbf{C1:} The local compactness $2m/r < 1$ to avoid singularities within the stellar configuration.

\textbf{C2:} Positive energy density ($\rho \geq 0$) and positive pressures ($P \geq 0$ and $P_{\perp} \geq 0$) . 

\textbf{C3:} Negative density gradient ($\rho^{\prime} < 0$) and negative pressure gradients ($P^{\prime} < 0$ and $P_{\perp}^{\prime} < 0$).

\textbf{C4:} Causality conditions on the radial and tangential speeds of sound ($0 < v_{s}^{2} \leq 1$ and $0 < v_{s\perp}^{2} \leq 1$).

\textbf{C5:} The restriction on the adiabatic index ($\Gamma = \frac{\rho + P}{P}v_{s}^{2} \geq \frac{4}{3}$) for stability against radial pulsations.

\textbf{C6:} The Harrison-Zeldovich-Novikov stability condition: $\mathrm{d}m_{b}(\rho_{c})/\mathrm{d}\rho_{c} > 0$.

\textbf{C7:} The cracking instability against local density perturbations~\cite{Herrera1992,DiPriscoHerreraVarela1997,AbreuHernandezNunez2007b}.

\textbf{C8:} The adiabatic convective stability condition ($\rho^{\prime\prime} \leq 0$).

In previous studies \cite{HernandezSuarezurangoNunez2021, suarez2023physical}, nine criteria were established to determine the physical acceptability of models. In this work, we relax the condition that requires the trace of the energy-momentum tensor to be positive. This condition is relevant for systems of non-interacting particles but does not necessarily apply to systems with strong interactions \cite{landau1975classical, Zeldovich:1961sbr}.

\begin{figure}[h]
\centering
\hspace{1.1cm}\includegraphics[width=2.55in]{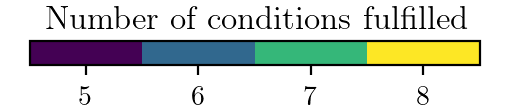}\\
\includegraphics[width=2.55in]{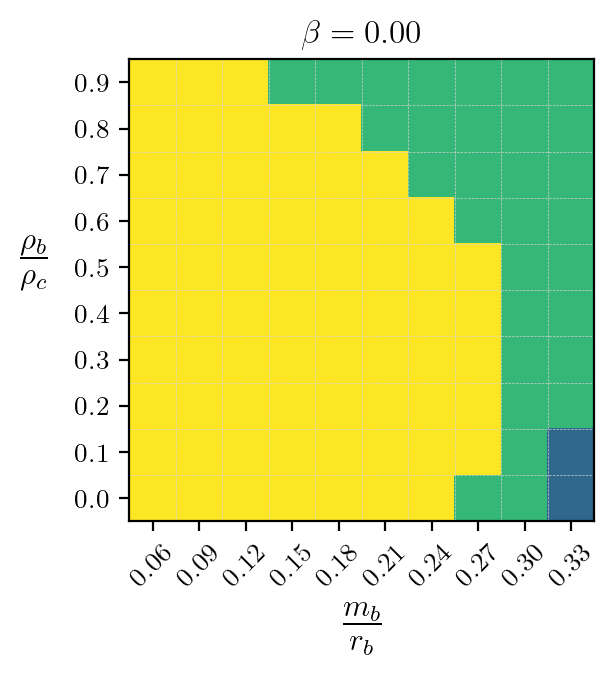}
\includegraphics[width=2.14in]{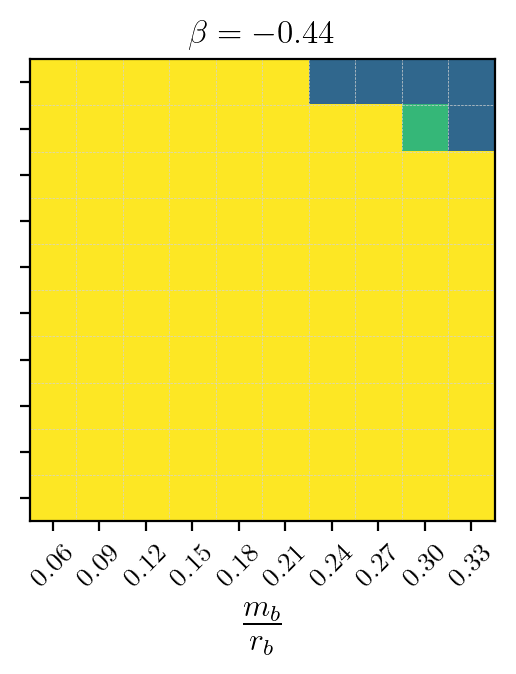}
\includegraphics[width=2.14in]{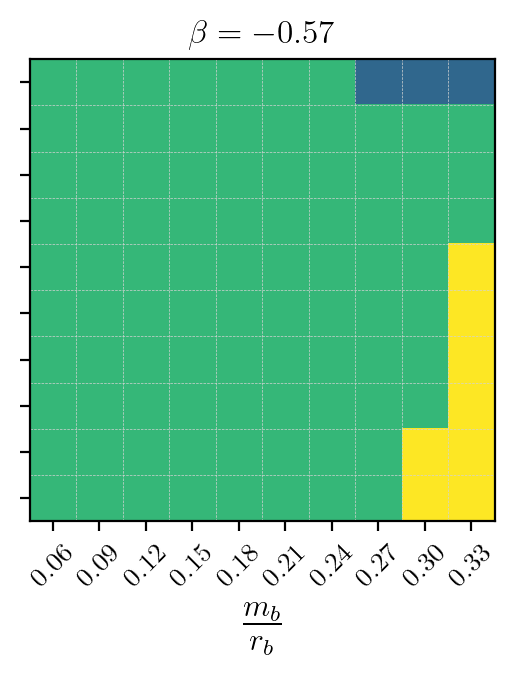}
\caption{Number of conditions fulfilled by stellar models with ($m_{b}/r_{b}$, $\rho_{b}/\rho_{c}$) as input parameters, for $\beta = 0.00$ (left plot), $\beta = -0.44$ (middle plot) and $\beta = -0.57$ (right plot). Physically acceptable models -fulfil all eight conditions- are painted yellow. Isotropic models ($\beta = 0.00$) become physically acceptable when local pressure anisotropy is considered.}
\label{ParSpac044}
\end{figure}

Assuming \(\tilde{\rho} = \beta \rho\) with \(\beta\) constant, integrating equations (\ref{HydAni2}) and (\ref{Massgradient}) requires only the density profile. Therefore, we adopt the Tolman VII density profile, given by \(\rho(r) = \rho_c(1 - \alpha r^2)\).  This profile was  previously used in \cite{suarez2023physical} to construct physically acceptable models based on NICER radius estimates for the pulsar PSR J0740+6620. In this context, the central density \(\rho_c\) and the constant \(\alpha\) (which is directly related to the surface density via \(\rho_b / \rho_c = 1 - \alpha r_b^2\)) are treated as free parameters.
\begin{table}[h]
\caption{This compilation table includes inferred values for equatorial radius ($R_{e}$ in kilometers), total mass ($m_{b}$ in solar masses), and compactness ($Gm_{b}/R_{e}c^{2}$) derived from NICER data for PSR J0030+0451 and PSR J0740+6620. All listed compactness values fall within the physically acceptable model region in the middle panel of Figure \ref{ParSpac044}, covering a broader range of $\rho_{b}/\rho_{c}$ than the isotropic case shown in the left panel. Note that the compactness of PSR J0740+6620 for Riley (2021) was indirectly calculated in this work by combining the upper and lower bounds of $m_{b}$ and $R_{e}$.} 
\centering
\begin{tabular}{|c | c c c |}
\hline
PSR J0030+0451 & $R_{e}\,[km]$ & $m_{b}\,[M_{\odot}]$ & $Gm_{b}/R_{e}c^{2}$  \\  \hline 
Miller (2019) \cite{miller2019psr} & $13.02^{+1.24}_{-1.06}$ & $1.44^{+0.15}_{-0.14}$ & $0.163^{+0.008}_{-0.009}$  \\ 
Riley (2019) \cite{riley2019nicer} & $12.71^{+1.14}_{-1.19}$ &  $1.34^{+0.15}_{-0.16}$  & $0.156^{+0.008}_{-0.010}$  \\
Vinciguerra (2024) \cite{vinciguerra2024updated} & $11.71^{+0.88}_{-0.83}$ &  $1.40^{+0.13}_{-0.12}$ & $0.177^{+0.006}_{-0.007}$   \\
Vinciguerra (2024) \cite{vinciguerra2024updated} & $14.44^{+0.88}_{-1.05}$ &  $1.70^{+0.18}_{-0.19}$  & $0.179^{+0.011}_{-0.022}$       \\ [0.5ex] \hline\hline

PSR J0740+6620 & $R_{e}\,[km]$ & $m_{b}\,[M_{\odot}]$ & $Gm_{b}/R_{e}c^{2}$  \\   \hline
Miller (2021) \cite{miller2021radius} & $13.71^{+2.62}_{-1.50}$ & $2.08\pm 0.07$ & $0.222^{+0.027}_{-0.035}$  \\ 
Riley (2021) \cite{riley2021nicer} & $12.39^{+1.30}_{-0.98}$ &  $2.07\pm0.07$ & $0.247^{+0.030}_{-0.031}$   \\
Salmi (2022) \cite{salmi2022radius} & $12.88^{+1.25}_{-0.95}$ &  $2.075\pm0.067$  & $0.238^{+0.018}_{-0.021}$  \\
Salmi (2024) \cite{salmi2024radius}  & $12.49^{+1.28}_{-0.88}$ &  $2.073\pm0.069$ & $0.245^{+0.017}_{-0.022}$  \\
Dittmann (2024) \cite{dittmann2024more} & $12.92^{+2.09}_{-1.13}$ & $2.08\pm0.07$ & $0.236^{+0.021}_{-0.032}$ \\ [0.5ex] 
\hline
\end{tabular} 
\label{NICERCompactness}
\end{table}

Once the total compactness \(\left(m_b / r_b\right)\), the surface-to-central density ratio \(\left(\rho_b / \rho_c\right)\), and the parameter \(\beta\) are specified, stellar models are obtained by numerically integrating equation (\ref{HydAni2}).  

Figure \ref{ParSpac044} illustrates the parameter space by varying typical values of \(m_b / r_b\) and \(\rho_b / \rho_c\). Each square in the figure represents a stellar model defined by these input parameters, with the color code indicating the number of physical acceptability conditions met by each configuration.   Notably, anisotropy enhances physical acceptability. In particular, isotropic models that fail to satisfy the causality condition (\(\mathbf{C4}\)) become acceptable when local pressure anisotropy is introduced. This suggests that pressure anisotropy allows for more compact, physically viable models.

\begin{figure}[h]
\centering
\includegraphics[width=6in]{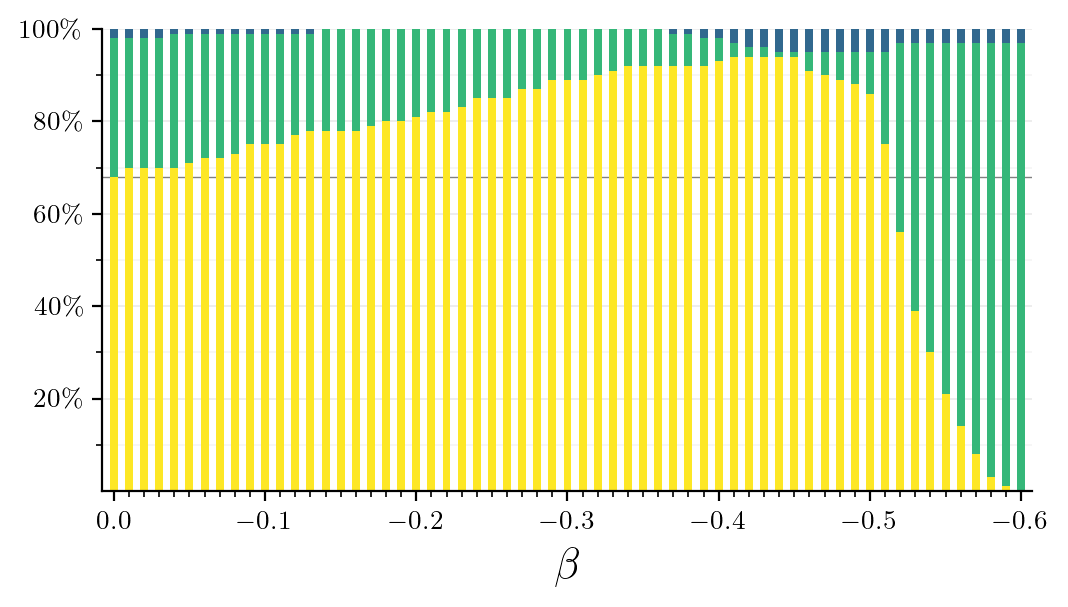}
\caption{Cumulative percentage of the number of physical acceptability conditions fulfilled in each parameter space for values of $\beta$ between $0.0$ and $-0.6$. The line in the background marks the percentage of acceptable models for the isotropic case ($\beta = 0.00$). The percentage of physically acceptable models (in yellow) increases until the maximum is $\beta = 0.41$ to $\beta = 0.45$. When $\beta \leq -0.52$, anisotropy does not improve physical acceptability. No physically acceptable models exist for $\beta \leq -0.60$.}
\label{AcParSpacBeta}
\end{figure}

This trend is illustrated in Figure \ref{AcParSpacBeta}, which displays the cumulative percentage of physical acceptability conditions satisfied across the parameter space for values of \(\beta\) ranging from $0.0$ to $-0.6$. As the magnitude of \(\beta\) increases, the proportion of physically acceptable models also rises, peaking for \(\beta\) values between -0.41 and -0.45. However, for \(\beta \leq - 0.52\), anisotropy no longer enhances acceptability, and for \(\beta \leq -0.60\), no models remain physically acceptable. This indicates that only moderate anisotropy improves the acceptability of isotropic configurations, consistent with previous studies \cite{suarez2022acceptability,suarez2023physical}.  

Table \ref{NICERCompactness} presents NICER-inferred compactness values for the pulsars PSR J0030+0451 and PSR J0740+6620 (see the cited references for details on the methodology and assumptions used). The possible compactness ranges are [$0.146$, $0.190$] for PSR J0030+0451 and [$0. 87$, $0.277$] for PSR J0740+6620. Within these ranges, the parameter space for \(\beta = -0.44\) yields more acceptable models than the isotropic case, as shown in Figure \ref{ParSpac044}. This suggests that anisotropic models provide a more realistic description of these observed pulsars.
\section{Dissipative case: Temperature as pressure contribution}
To show the impact of dissipation, in this section, we will use the Israel-Stewart heat flow transport equation, $\left(57\right)$ in reference \cite{HerreraEtal2014}
\begin{equation}
\tau h^\nu_\mu \mathcal{F}^\nu_{;\beta}V^\beta+\mathcal{F}^\mu=-\kappa h^{\mu\nu}\left (T_{,\nu}+Ta_\nu  \right)-\frac{1}{2}\kappa T^2\left ( \frac{\tau V^\alpha}{\kappa T^2} \right)_{;\alpha}\mathcal{F}^\mu ,
\end{equation}
where $\kappa$ is the thermal conductivity and $\tau$ is the relaxation time, to obtain the general expression of the TOV equation for a spherical, dissipative and anisotropic fluid as follows
\begin{equation}
    \left(\mathbf{P}-\frac{\kappa}{\tau}T\right)_{,\beta} h^\beta_\alpha = -\left(\rho+\mathbf{P}-\frac{\kappa}{\tau}T\right)a_\alpha - \Pi_{\beta;\mu}^\mu h^\beta_\alpha
-\sigma_{\alpha\beta}\mathcal{F}^\beta 
+\left\{\frac{1}{\tau}+\frac{1}{2} \left[\ln \left(\frac{\eta \tau}{\kappa T^2}\right)\right]^{\bullet} -\frac{5}{6}\Theta \right\}\mathcal{F}_\alpha\,, 
\end{equation}
here $\eta$ is a constant of integration with inverse length units.  Notice that thermal effects influence both the pressure gradient and the pressure itself. In this scenario, dissipation is treated as an additional contribution to the radial pressure, effectively transforming any dissipative anisotropic model into an equivalent non-dissipative isotropic configuration. This provides a simpler and more intuitive approach to describing compact objects.

Assuming  $\frac{\kappa}{\tau}=const$, we have
\begin{equation}
    \bar{P}_{,\beta}h^\beta_\alpha=-\left(\bar{\rho}+\bar{P}\right)a_\alpha+\bar{\sigma} \mathcal{F}_\alpha\label{TOVDis}
\end{equation}
 where
\begin{equation}    
\bar{\rho} = \rho + \tilde{\rho}\,, \qquad \bar{P} = P -\frac{\kappa}{\tau}T \qquad \textrm{and} \qquad 
\bar{\sigma} = \frac{1}{\tau}-\frac{\dot{T}}{AT}-\frac{2}{3}\sigma-\frac{5}{6}\Theta\,.\label{barPDis}
\end{equation}
Again, for spherical symmetry, we have
\begin{equation}
    a_\alpha=a_1 K_\alpha \qquad \textrm{and} \qquad \mathcal{F}_\alpha= \mathcal{F} K_\alpha \, .
\end{equation} 
Thus, equation (\ref{TOVDis}) can be rewritten as
\begin{equation}
    \bar{P}_{,\beta}h^\beta_\alpha=\,-\left(\bar{\rho}-\frac{\bar{\sigma}\mathcal{F}}{a_1}+\bar{P}\right)a_\alpha \label{TOVDis2}\,.
\end{equation}
Anisotropy influences energy density, while temperature influences radial pressure. 
This work will consider only shear-free fluids to simplify the dissipative models. In this case 
\begin{equation}
    \sigma_1=\sigma_2\,\, \Rightarrow \, R= f(r) B\label{SFcond}\,,
\end{equation}
without loss of generality, we can always choose a particular coordinate system where $f(r)=r$, and furthermore,
 \begin{equation}
     \bar{\sigma} = 0  \quad \Rightarrow\quad \frac{\dot{T}}{T} = \frac{A}{\tau} - \frac{5\dot{R}}{2R}  \quad \Rightarrow \quad T(t,r)=\frac{T_0(r)}{R^{\frac{5}{2}}}e^{\int \frac{Adt}{\tau}}\label{CT1}\,.
 \end{equation}
Now, taking into account (\ref{CT1}), equation (\ref{TOVDis2}) can be written as follows
\begin{equation}
    \bar{P}_{,\beta}h^\beta_\alpha=\,-\left(\bar{\rho}+\bar{P}\right)a_\alpha \label{TOVDis3}\,.
\end{equation}
This equation has the same form as the TOV equation for a static perfect fluid. This allows us to find anisotropic and dissipative models inspired by perfect fluid models.
In the following section, we shall develop two simple models to illustrate this effect.

\subsection{Florides-Like Model}
Relativistic static matter configurations with vanishing radial stresses date back to the work of G. Lema{\^i}tre \cite{Lemaitre1933}, and P.S. Florides \cite{Florides1974}. Modelling such configurations, where only tangential forces are present, has led to valuable understandings of such matter distributions, including developing non-local equations of state \cite{HernandezNunez2004} and the analysis of the complexity and simplicity of relativistic matter structures.

In this section, we adopt this type of Florides-like model with $P\neq 0$ but having $\bar{P} = 0$, implying that $P = \frac{\kappa}{\tau}T$ (from equation \ref{TOVDis2}), which provides a simplified model for analysis to integrate TOV equation. Thus, we have    
\begin{equation}
\bar{\rho}a_\alpha=0 \,,\label{TOVDis4}
\end{equation}
and to recreate the most similar situation to the gravitational collapse of an Oppenheimer-Snyder dust cloud, we will opt for the geodesic and isotropic pressure case, that is
 \begin{eqnarray}
     a_\alpha=0 \quad \Rightarrow \quad \frac{A^{\prime}}{A} = 0 \quad \Rightarrow \quad  A(t,r)=1\,.\label{GeodC1}
 \end{eqnarray}
 Moreover, due to (\ref{GeodC1}) then $a_1 = 0$ and consequently the set of equations (\ref{ecR1})-(\ref{ecR8}) can be written as follows
\begin{align}
Y_0&=-3\sigma_{1}^{\bullet}-3\sigma_{1}^2 \, , \label{ecf1} \\
Y_1\,&=0 \, , \label{ecf2}  \\
Z\,&=\,-\sigma_{1}^\dag \, , \label{ecf3} \\
X_0\, &=\, 3\sigma_{1}^2+\frac{3}{R^2}-3J_1^{2}-2 X_1\, , \label{ecf4} \\
X_1\,&=\, J_1^\dag +\frac{1}{R^2}\, .\label{ecf5} 
\end{align}
Notice that $Y_{1} = 0$ arises from subtracting (\ref{ecR1}) from (\ref{ecR2}) and considering the geodesic and shear-free (\ref{GeodC1}) condition.

On the other hand, taking into account (\ref{ecf2}), the isotropic pressure condition implies $X_1=0$ and thus, from (\ref{ecf5}) we obtain a differential equation for the metric function $R(t,r)$, namely
\begin{equation}
    \frac{R^{\prime \prime}}{R}+\frac{R^\prime }{rR}-2\left(\frac{R^\prime}{R}\right)^2+\frac{1}{r^2}=0\, ,
 \end{equation}
Whose solution, satisfying the regularity condition at $r\rightarrow 0$, is
\begin{equation}
   R(r,t)=\frac{ r}{\tilde R(t)r^2+1}\,,  \label{ecRf} 
 \end{equation}
and 
\begin{equation}
   B(r,t)=\frac{1 }{\tilde{R}(t)r^2+1} \,.  \label{ecBf} 
 \end{equation}

Now, by using (\ref{GeodC1}), (\ref{ecRf}) and (\ref{ecBf}), from the equations (\ref{ecf1})-(\ref{ecf4}) we find

\begin{align}
Y_0&=\frac{3r^2 \left(r^2\dot{\tilde{R}}(t)+1\right)\ddot{\tilde{R}}(t)-6 r^4 \dot{\tilde{R}}(t)^2}{\left(r^2 \tilde{R}(t)+1\right)^2}\, , \label{ecff1} \\
Y_1\,&=0 \, , \label{ecff2}  \\
Z\,&=\,\frac{2 r \dot{\tilde{R}}(t)}{r^2  \tilde{R}(t)+1} \, , \label{ecff3} \\
X_0\, &=\,  3\left(\frac{ r^2 \dot{\tilde{R}}(t)}{1 + r^2 \tilde{R}(t)^2}\right)^2+12 \tilde{R}(t)\,,
\label{ecff4} \\
X_1\, &=0\,.
\end{align}
Finally, for the temperature, from (\ref{CT1}), we get
\begin{equation}
    T(t,r)=\frac{T_0(r) (\tilde R(t) r^2+1)^{\frac{5}{2}}e^{\frac{t}{\tau}}}{r^{\frac{5}{2}}}
\end{equation}

 \subsection{Schwarzschild-Like Model}
 We have shown that the most simple and ``pedagogic'' spherical matter solution --$\rho = \mbox{const}$--  is very restricted and mostly unphysical~\cite{OspinoEtal2018}.  Despite its physical inconsistency --it models a fluid with an infinite sound speed-- its simplicity is of a pedagogical value in illustrating the methods used in solving physical systems in different (static \& dynamic) interesting scenarios \cite{BowersLiang1974, HerreraNunez1989, Wyman1949, BonnorFaulkes1967, MisraSrivastava1973, Ponce1986, HerreraNunez1987, MaharajMaartens1989,  NunezRueda2007}. Thus, in this example, we will also assume 
\begin{equation}
    \bar \rho = \bar \rho_0(t).
\end{equation}
 After  integration of equation (\ref{TOVDis3}) we get
 \begin{equation}
     A(r,t)=\tilde{A}(t)\left (\bar \rho_0 (t)+\bar P(r,t)\right )\label{ecAS},
 \end{equation}
where $\tilde{A}(t)$ is an arbitrary integration function. It is worth noting that in this case the condition $X_1=0$ does not lead to a perfect fluid, therefore, using this condition we can find anisotropic and dissipative solutions. In other words, let us assume again that the metric functions $R$ and $B$ are given by (\ref{ecRf}) and (\ref{ecBf})  respectively. To determine the metric function $A(r,t)$ and obtain anisotropic and dissipative models close to the Schwarzschild interior solution \cite{SchwInt1916}, we choose the function $\bar P$ as follows:
\begin{equation}
\bar P(r,t)=\bar P_0(t)\sqrt{1-\tilde{R}(t)r^2} \label{barPS},
\end{equation}
and from (\ref{ecAS}) we get
 \begin{equation}
     A(r,t)=\alpha_1(t)-\alpha_2(t)\sqrt{1-\tilde{R}(t)r^2},
 \end{equation}
with $\alpha_1(t)=\tilde{A}(t)\bar \rho_0 (t)$ and $\alpha_2(t)=\tilde{A}(t)\bar P_0(t)$. Once the metric functions are known, from the equations (\ref{ecR1})-(\ref{ecR8}), we find the expressions for the physical variables

\begin{align}
Y_0\,&=\, \frac{\tilde{R}(t)  \alpha_2(t) \left(1 + r^2 \tilde{R}(t)\right)(3-r^2 \tilde{R}(t))}{\left(1-r^2 \tilde{R}(t)\right)^{3/2} A(r,t)}-\frac{3}{2A^2(r,t)(1+r^2\tilde{R})}\left( -2r^2\ddot{\tilde{R}}\,+ \right . \label{ecSch1}\nonumber \\
&\,\quad \left. r^4\dot{\tilde{R}}^2(t)\left(\frac{\alpha_2(t)}{A(r,t)\sqrt{1-r^2\tilde{R}}}+\frac{4}{1+r^2\tilde{R}(t)}\right)+\frac{2\dot{\tilde{R}}r^2}{A(r,t)}\left (\dot{\alpha}_1-\sqrt{1-r^2 \tilde{R}}\dot{\alpha}_2(t)\right) \right )\,,\\
Y_1\, &=\,-\frac{\tilde{R}^2(t) r^2 \left(1 + r^2 \tilde{R}(t)\right) \left(3r^2 \tilde{R}(t) -5 \right)\alpha_2(t)}{\left(1-r^2 \tilde{R}(t)\right)^{3/2} A(r,t)}\,, \label{ecSch2} \\
Z\, &= \, \left(\frac{2 A(r,t)}{1 + r^2 \tilde{R}(t)}-\frac{r^2 \alpha_2(t) \tilde{R}(t)}{\sqrt{1 - r^2 \tilde{R}(t)}}\right)\frac{r\dot{\tilde{R}}(t)}{A^2(r,t)}\, ,\label{ecSch3}\\
X_0\, &= \, 12 \tilde{R}(t) + \frac{3}{A^2(r,t)}\left(\frac{ r^2 \dot{\tilde{R}}(t)}{1 + r^2 \tilde{R}(t)^2}\right)^2 \label{ecSch4}\,, \\
X_1\, &= \, 0 .
\end{align}
Finally, combining the equations (\ref{barPDis}), (\ref{barPS}), (\ref{ecSch1}) and (\ref{ecSch4}),  we can obtain the  temperature of the configuration:
\begin{eqnarray}
   \frac{\kappa}{\tau} T&=&\frac{2Y_0-X_0-2Y_1}{24\pi}-\bar{P}\,.
\end{eqnarray}
\section{Final remarks}
Using a tetrad formalism that expresses Einstein equations through structure scalars, we reinterpret two fundamental aspects of spherically symmetric relativistic stellar configurations: 
\begin{itemize}
	\item the impact of local pressure anisotropy on energy density, and 
	\item the effect of dissipation on radial pressure.
\end{itemize}  

When analyzing pressure anisotropy in relativistic stellar structures, we find that modifications in energy density offer a new perspective on these matter configurations. If the proportionality factor \( \beta \), which relates energy density to anisotropy, is constant (\( \beta < 1 \)), anisotropy directly modifies the energy density. In more general cases, where \( \beta \) varies with the radial coordinate, this factor provides localized information, allowing for the modeling of anisotropic equations of state in a more detailed and flexible manner.  

To validate this reinterpretation, we derive models based on a particular given density profile and demonstrate that many of them satisfy physical acceptability conditions. Moreover, introducing anisotropy improves the physical acceptability of stellar configurations. We compare our results with compact object data from NICER and find that the obtained models can accurately describe astrophysical objects such as PSR J0740+6620 and PSR J0030+0451.  

Dissipation is treated as an additional contribution to the radial pressure, allowing any  dissipative anisotropic model  to be transformed into an  equivalent non-dissipative isotropic configuration --providing the simplest description of  compact objects. For  anisotropic dissipative configurations, we present two models:
\begin{itemize}  
	\item A  Florides-like model, characterized by \( \bar{P} = 0 \) and a  shear-free condition, resembling the  gravitational collapse of an Oppenheimer-Snyder dust cloud.  
	\item A  Schwarzschild-like model, with an  effective homogeneous density  and a  well-defined pressure profile  given by equation (\ref{barPS}).  
\end{itemize}
Our approach clarifies the role of  anisotropy and thermal effects, simplifying the resolution of  Einstein's field equations  in  dissipative and anisotropic systems.
 
\section*{Acknowledgments}
L.A.N. and D.S.U. acknowledge the financial support of the Vicerrector\'ia de Investigaci\'on y Extensi\'on de la Universidad Industrial de  Santander and Universidad de Salamanca through the research mobility programs. L.A.N. also thanks the hospitality of the Departamento de Matem\'aticas Aplicadas, Universidad de Salamanca. J.O.  acknowledge financial support from Ministerio de Ciencia, Innovaci\'on y Universidades (grant PID2021-122938NB-100) and Junta de Castilla y Le\'on  (grant SA097P24). J.O. acknowledges the hospitality of the School of Physics of t e Industrial University of Santander, Bucaramanga, Colombia.

\bibliographystyle{unsrt}
\bibliography{AnisotropyDensity}



\end{document}